\documentclass[runningheads]{llncs}
\usepackage[T1]{fontenc}
\usepackage{graphicx,verbatim}
\usepackage{amsmath}
\usepackage{amssymb}
\usepackage{multicol}
\usepackage{multirow}
\usepackage{caption} 
\usepackage{float}
\usepackage{hyperref}
\usepackage{cleveref}
\DeclareMathOperator{\Loss}{\mathcal{L}}
\usepackage{color}

\begin{document}
\title{D2Diff: A Dual-Domain Diffusion Model for Accurate Multi-Contrast MRI Synthesis}
\titlerunning{D2Diff: A Dual-Domain Diffusion Model}
%



\author{Sanuwani Dayarathna\inst{1} \and
Himashi Peiris\inst{1} \and
Kh Tohidul Islam\inst{2} \and
Tien-Tsin Wong\inst{1} \and
Zhaolin Chen\inst{1,2}}
\authorrunning{S. Dayarathna et al.}
%
\institute {Department of Data Science and AI, Faculty of IT, Monash University, Melbourne, Australia. \and
Monash Biomedical Imaging, Monash University, Melbourne, Australia.
\email{\{Sanuwani.Hewamunasinghe,Himashi.Peiris,KhTohidul.Islam,\\TT.Wong,Zhaolin.Chen\}@monash.edu}}

\maketitle              
\begin{abstract}

Multi-contrast MRI synthesis is inherently challenging due to the complex and nonlinear relationships among different contrasts. Each MRI contrast highlights unique tissue properties, but their complementary information is difficult to exploit due to variations in intensity distributions and contrast-specific textures. Existing methods for multi-contrast MRI synthesis primarily utilize spatial domain features, which capture localized anatomical structures but struggle to model global intensity variations and distributed patterns. Conversely, frequency-domain features provide structured inter-contrast correlations but lack spatial precision, limiting their ability to retain finer details. To address this, we propose a dual-domain learning framework that integrates spatial and frequency domain information across multiple MRI contrasts for enhanced synthesis. Our method employs two mutually trained denoising networks, one conditioned on spatial domain and the other on frequency domain contrast features through a shared critic network. Additionally, an uncertainty-driven mask loss directs the model’s focus toward more critical regions, further improving synthesis accuracy. Extensive experiments show that our method outperforms state-of-the-art (SOTA) baselines, and the downstream segmentation performance highlights the diagnostic value of the synthetic results. Code and model hyperparameters are available at \href{https://github.com/sanuwanihewa/D2Diff}{https://github.com/sanuwanihewa/D2Diff}


\end{abstract}
\section{Introduction}

 Magnetic Resonance Imaging (MRI) offers detailed anatomical and pathological insights through images of multiple contrasts \cite{Chen2022}. However, acquiring multiple MRI contrasts poses significant challenges, including high imaging cost, prolonged scanning times, and potential safety concerns related to gadolinium-based contrast agents \cite{Dayarathna2024,Lohrke2016}. Medical image synthesis provides a powerful approach to address these challenges by reconstructing missing or corrupted image contrasts from available contrasts \cite{Yu2020}. However, synthesizing high-fidelity multi-contrast images remains challenging due to the complex, nonlinear, and often obscured relationships among contrasts, driven by intensity inconsistencies and modality-specific textures \cite{Dayarathna2024}. Therefore, capturing and aligning these intricate cross-contrast relationships is critical for an accurate image synthesis model.
 
 Most of the existing methods \cite{Dayarathna2024,Zhou2020,Zhan2022} focus on fusing features from multiple contrasts, leveraging latent-level operations or hierarchical representations to model cross-contrast dependencies. Despite these advances, most approaches \cite{Sharma2020,Zhou2020} rely heavily on rigid spatial domain representations and fusion strategies, which struggle to fully capture the complementary and distributed relationships across contrasts. Although spatial domain features excel at encoding localized structures and anatomical integrity, they often struggle to disentangle discerning intensity variations and overlapping distributions \cite{Lou2024,Ding2023}, particularly in scenarios with significant heterogeneity across contrasts such as brain lesions\cite{Dayarathna2024}.

To address these limitations, we propose a dual-domain learning framework, D2Diff, for multi-contrast MRI synthesis, which employs two denoising networks that are mutually trained together. The first network is guided by frequency-domain representations \cite{Lou2024,Ding2023} and captures structured inter-contrast correlations such as global intensity shifts and distributed intensity variations. Simultaneously, the second network is guided by spatial-domain features and ensures high-resolution, pixel-level detail fidelity. These networks are trained collaboratively through a shared critic network, which ensures adversarial consistency. Using a novel uncertainty-aware mask loss, the shared critic facilitates uncertainty estimation, guiding the synthesis process to focus on critical regions. By leveraging the complementary strengths of spatial and frequency domains, our framework effectively aligns complex cross-contrast correlations, providing a robust and accurate multi-contrast MRI synthesis. In summary, \textbf{our main contributions are}:
\textbf{(1) }A dual-domain diffusion framework, simultaneously guided by multi-contrast MRI features in both frequency and spatial domains, and jointly trained using a shared critic network. \textbf{(2) } A multi-scale frequency feature integration module for adaptive inter-contrast feature combination to preserve subtle contrast-specific details. \textbf{(3) }A novel uncertainty-aware mask loss to enhance uncertainty-driven learning. \textbf{(4) }Comprehensive experiments confirm superior synthesis quality and further validation through downstream segmentation tasks.

\section{Method}

\begin{figure*}[htb]
\includegraphics[width=\textwidth]{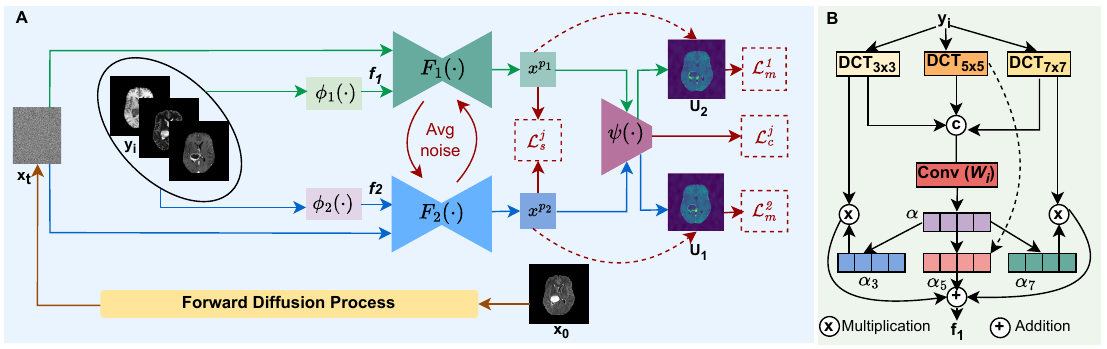}
\caption{Network architecture. \textbf{A}: Overall Architecture with frequency ($\phi_1$) and spatial guidance ($\phi_2$), \textbf{B}: Multi-scale adaptive frequency aggregation in $\phi_1$. 
} \label{fig:framework}
\end{figure*}

\paragraph{\textbf{Problem Formulation.}}Let $\mathcal{X}={(\vec{X}_k, \vec{Y}_k)}_{k=1}^m$, be a set of m co-registered MRI contrast image pairs, where $\vec{x_k}$  denotes the target contrast to be synthesized, and $\vec{y_k} = \{\vec{y_{k,i}}\}_{i=1}^n$, represents n source contrasts used as conditional inputs to generate the target contrast. We denote the denoising networks as $\mathcal{H}_j; j \in \{1,2\}$ where encoder-decoder $\mathcal{F}_j$, and dual-domain feature extraction $\phi_j$ are their functional decompositions as,
\begin{equation}
 f_j = \phi_j (\theta_j;\vec{Y}_k);
 \quad
 \mathcal{H}_j(\Theta^j; \mathcal{X}) = \mathcal{F}_j (\phi_j (\theta_j;\vec{Y}_k),\vec{X}_k) \quad j= 1 ~\text{or}~ 2.
   \label{eqn:equation1}
\end{equation}


\paragraph{\textbf{Dual Domain Diffusion Model.}} Fig. \ref{fig:framework}(A) provides an overview of the D2Diff pipeline, which employs two denoising networks that collaboratively learn using frequency and spatial domain features from multi-contrast MRI. Diffusion models consist of two main processes: forward and reverse process \cite{Ho2020}. In the forward process, random Gaussian noise is progressively added to the target MRI contrast ($\vec{x}_0$) to be synthesized as,
\begin{equation}
 \begin{aligned}
  q(\vec{x}_\text{t} | \vec{x}_{\text{t}-1}) = \mathcal{N}  (\vec{x}_\text{t};\sqrt{1-\beta_\text{t}} \; \vec{x}_{\text{t}-1}, \beta_\text{t} \text{I} ),
   \label{eqn:equation2}
 \end{aligned}
\end{equation}
where $\beta_t$ is the noise variance schedule that is used to add noise to the data, $\mathcal{N}$ is the Gaussian distribution, and $I$ is the identity covariance matrix. Utilizing the Markov property of the diffusion process, the marginal distribution of $\vec{x}_\text{t}$ can be directly obtained as follows,
\begin{equation}
 \begin{aligned}
q(\vec{x}_\text{t}|\vec{x}_0) = \mathcal{N}  (\vec{x}_\text{t}; \sqrt{\bar{\alpha_\text{t}}} \; \vec{x}_0, (1-\bar{\alpha_\text{t}})\text{I} ),
   \label{eqn:equation3}
 \end{aligned}
\end{equation}
where $\alpha_\text{t} := 1-\beta_\text{t}$ and $\bar{\alpha_\text{t}} := \prod_{s=1}^{\text{t}} \alpha_s$. The reverse diffusion process estimates the posterior distribution $p_\theta (x_{t-1}|x_t,\vec{y}_i)$ to generate a realistic $x_0$ guided by conditional contrasts $\vec{y}_i$,
\begin{equation}
 \begin{aligned}
p_\theta (\vec{x}_{\text{t}-1}|\vec{x}_\text{t},\vec{y}_i) = \mathcal{N}  (\vec{x}_{\text{t}-1};\mu_\theta(\vec{x}_\text{t},\text{t}),\sigma^2_\text{t} \text{I} ),
   \label{eqn:equation4}
 \end{aligned}
\end{equation}
where $\mu_\theta(\vec{x}_\text{t},\text{t})$ is the mean and $\sigma^2_t$ is the variance of the denoising network parameterized by $\theta$. The noisy target contrast serves as input for both frequency- and spatial-guided synthesis models. Each denoising network then independently performs the reverse diffusion process on the perturbed data, leveraging multi-contrast features in their respective domains to approximate the posterior distribution parameterized \cite{Xiao2021} as follows,
\begin{equation}
 \begin{aligned}
p_\theta(\vec{x}_{\text{t}-1}|\vec{x}_\text{t},\vec{y}_i) := q(\vec{x}_{\text{t}-1}| \vec{x}_\text{t},\tilde{\vec{x}}_0 = H_j(\vec{x}_\text{t},\vec{y_i},z,\text{t})).
   \label{eqn:equation5}
 \end{aligned}
\end{equation}
\paragraph{\textbf{Frequency domain learning.}}
Multi-contrast MRI exhibits significant variations in intensity and resolution across contrasts. The frequency domain structures spatial patterns into different frequency components, enabling better separation of global and local intensity variations\cite{Lou2024}.  
To leverage this, we apply Discrete Cosine Transform (DCT) \cite{Ahmed} to convert spatial variations into frequency representations, aligning non-linear intensity differences. DCT employs real-valued cosine functions rather than complex exponentials, allowing efficient decomposition of MRI images into frequency components \cite{Wei2024,Ziashahabi}. By transforming multi-contrast images into frequency coefficients, DCT effectively distributes intensity variability across distinct bands, enhancing feature consistency for synthesis. To achieve this, we used DCT with three different kernel sizes ($\text{k}$) in $\phi_1(\cdot)$, allowing the extraction of multi-scale frequency features as follows.
\begin{equation}
 \begin{aligned}
  h_{k,\vec{y}_i} = \text{DCT}_{k \times k}(\vec{y}_i), \quad \text{for } k \in \{3,5,7\}
   \label{eqn:equation6}
 \end{aligned}
\end{equation}
where $h_{k,y_i}$ is the the extracted frequency contrasts of each $y_i$.

To optimize frequency feature weighting, we introduce a novel adaptive feature aggregation using a lightweight attention mechanism that assigns importance scores via a learnable attention module and a convolutional layer. This refines and projects the combined representation into a common space. The adaptive frequency fusion (Fig. 1B) selectively emphasizes relevant frequency-specific information across MRI contrasts as follows,
\begin{equation}
 \begin{aligned}
 f_1 =  \Big[ W_i \cdot \sum_{k \in \{3,5,7\}} \left\langle \alpha_{k} , h_{k,\vec{y}_i}\right\rangle \Big] _{i \in \{1,n\}}
   \label{eqn:equation7}
 \end{aligned}
\end{equation}
where \( W_i \) is a learnable convolutional transformation layer and $ \alpha_{k}$ represents the adaptive feature combination process as shown in Fig. 1 (B) where a softmax-based attention ($\alpha$) mechanism is used to assign dynamic weights ($\alpha_3$, $\alpha_5$, $\alpha_7$) to determine how much influence each contrast's frequency features should have in the final representation. The weighted sum of these features forms the fused representation $f_1$, to guide the first denoising network.
\paragraph{\textbf{Spatial  domain learning.}}
To preserve fine anatomical details, the second denoising network is guided by spatial features from multi-contrast inputs in $\phi_2(\cdot)$. This enhances structural correlations, capturing finer details like edges and tissue boundaries to aid the denoising process as follows,
\begin{equation}
 \begin{aligned}
 f_2 =  [R_i(\vec{y}_i) ]_{i \in \{1,n\}}
   \label{eqn:equation8}
 \end{aligned}
\end{equation}
where $R_i$ consists of separate residual blocks for each input contrast, which consist of a convolutional layer followed by a Group Normalization, ReLU activation. 

Both denoising networks $\mathcal{F}_1$, $\mathcal{F}_2$ employs U-Net-based architecture as in \cite{Xiao2021} while sinusoidal positional embeddings \cite{Ho2020} encode the timestep t with $z$ serving as the latent vector for conditioning. 
\begin{equation}
 \begin{aligned}
 \mathcal{H}^{\Theta}_j(X) = \mathcal{F}_j(f_j,\text{t},z), ~j\in \{1,2\}
   \label{eqn:equation9}
 \end{aligned}
\end{equation}

Alongside the denoising generators, we employ a shared time-dependent critic network  $\psi$ \cite{Xiao2021} to ensure collaborative training across them. $\psi$ distinguish between $\vec{x}_{\text{t}-1}$ and $\vec{x}_\text{t}$ by assessing if $\vec{x}_{\text{t}-1}$ is a plausible denoised version of $\vec{x}_\text{t}$ using the critic loss $\Loss_c^j$.
\begin{equation}
 \begin{aligned}
\Loss_c^j\,(\theta^j_\mathcal{H}; \mathcal{X}) = \mathbb{E}_{q(\vec{x}_\text{t}|\vec{x},\vec{y}_i), p_\theta(\vec{x}_{\text{t}-1}|\vec{x}_\text{t},\vec{y}_i)}  [-log (\psi^{\theta} (\vec{x}^{p_j}_{\text{t}-1},\vec{x}_\text{t},\text{t})) ]
   \label{eqn:equation10}
 \end{aligned}
\end{equation}
As each denoising network is trained on different feature domains of the same input contrasts, they can leverage the shared critic network to learn from each other, maintaining consistency in their predictions.
To train the critic network against the actual ground truths, the predicted outputs ($\vec{x}^{p_1}$ and $\vec{x}^{p_2}$) from each denoising network are used as follows,
\begin{equation}
 \begin{aligned}
  \Loss_{adv}^j (\theta_c; \mathcal{X}) = \mathbb{E}_{q (\vec{x}_{\text{t}} |\vec{x},\vec{y}_i)} [\mathbb{E}_{q(\vec{x}_{\text{t}-1}|\vec{x}_\text{t},\vec{y}_i)} \eta  [ log (\psi^\theta (\vec{x}_{\text{t}-1},\vec{x}_\text{t},\text{t}) ) ] \\ + (1-\eta) \mathbb{E}_{p_\theta (\vec{x}_{\text{t}-1}|\vec{x}_\text{t},\vec{y}_i)}  [log (1-\psi^\theta (\vec{x}^{p_j}_{\text{t}-1},\vec{x}_\text{t},\text{t}) ) ] ],
\label{eqn:equation11}
 \end{aligned}
\end{equation}
where $\eta = 0$ if $\vec{x}_{\text{t}-1}$ is predicted by a denoising network, and $\eta = 1$ if $\vec{x}_{\text{t}-1}$ is sampled from the actual target contrast distribution.

\paragraph{\textbf{Uncertainty aware mask loss.}}
We propose an uncertainty-aware mask loss that guides denoising networks to focus on high-uncertainty regions during the synthesis. This is achieved using spatial attention maps from the critic network, which identify reliable features from the target distribution. Specifically, we considered middle-layer features ($f_m$) from the critic network, which is sensitive to discriminative regions extracted via a sigmoid($\sigma$) layer and interpolated ($I$) to match the output contrast dimension ($dim$) as,
\begin{equation}
 \begin{aligned}
  U_j = I  [\sigma  [\psi^\theta (\vec{x}_{\text{t}-1}^{p_j},\vec{x}_\text{t},\text{t})_{f_m} ] , dim(\vec{x}) ]
\label{eqn:equation12}
 \end{aligned}
\end{equation}

To enhance mutual learning across networks, each denoising network leverages the attention maps of other's output contrast from the shared critic to align individual predictions. Then, the Binary cross-entropy logistic criteria (BCE) is employed to quantify discrepancies, encouraging consistent probability estimations and refining focus on critical regions using mask loss $\Loss_{m}^j$ as,
\begin{equation}
 \begin{aligned}
\Loss_{m}^j\,(\theta^j_\mathcal{H}; \mathcal{X}) =  \langle U_2,BCE (\vec{x}^{p_1}, \sigma (\vec{x}^{p_2}) ) \rangle+  \langle U_1, BCE (\vec{x}^{p_2}, \sigma (\vec{x}^{p_1}) ) \rangle
\label{eqn:equation13}
 \end{aligned}
\end{equation}

We also employed supervised loss $\Loss_{s}^j$ between individual predictions from each network and actual contrast as follows,
\begin{equation}
 \begin{aligned}
\Loss_{s}^j\,(\Theta^j_\mathcal{H}; \mathcal{X}) = \mathbb{E}_{(\vec{x},\vec{y}_i\epsilon \mathcal{X})}\left\| \vec{x}-\vec{x}^{p_j}\right\|_1
   \label{eqn:equation14}
 \end{aligned}
\end{equation}

Then, the two denoising networks are trained by minimizing the  objective,
\begin{equation}
 \begin{aligned}
 \Loss\,(\Theta^j_\mathcal{H};\mathcal{X}) = \sum_{j=1}^2 [\lambda_s \, \Loss_s^j \, (\Theta^j_\mathcal{H};\mathcal{X}) +\lambda_m \, \Loss_m^j \, (\Theta^j_\mathcal{H};\mathcal{X})+ \lambda_c \,  \Loss_c^j \, (\Theta^j_\mathcal{H};\mathcal{X}) ]
   \label{eqn:equation15}
 \end{aligned}
\end{equation}
where $\lambda_s, \lambda_m, \lambda_c > 0$ control the contribution of each loss component.
\paragraph{\textbf{Dual-domain consistency.}}
During the inference process, we start at timestep T with random Gaussian noise as $\vec{x}_t$ and iteratively refine through T number of sampling steps. At each step we derive $\text{t}-1^{th}$ sample using Markov property of forward process\cite{Ho2020} as follows,
\begin{equation}
 \begin{aligned}
q(\vec{x}_{\text{t}-1}|\vec{x}_\text{t},\vec{x}_0 = \vec{x}^{p_j}) = \mathcal{N} \Big(\vec{x}_{\text{t}-1};\tilde{\mu_\text{t}}(\vec{x}_\text{t},\vec{x}^{p_j}),\tilde{\beta_\text{t}}\text{I} \Big)
   \label{eqn:equation16}
 \end{aligned}
\end{equation}
where $\tilde{\mu_\text{t}}$ and  $\tilde{\beta_\text{t}}$ is the mean and variance of the distribution.

To ensure mutual learning between two denoising networks, we derive the average mean noise predictions across two networks using eq.\ref{eqn:equation3} and eq.\ref{eqn:equation4} as below,
\begin{equation}
 \begin{aligned}
\tilde{\mu_\text{t}}_{avg}(\vec{x}_\text{t},\vec{x}^{p_j}) := \frac{1}{2}\sum_{j=1}^{2} \Big[\frac{\sqrt{\bar{\alpha}_{\text{t}-1}}\:\tilde{\beta_\text{t}}}{1-\bar{\alpha}_{\text{t}}} \: \vec{x}^{p_j} \:+\: \frac{\sqrt{\alpha_\text{t}}\:(1-\bar{\alpha}_{\text{t}-1})}{1-\bar{\alpha}_{\text{t}}} \: \vec{x}_\text{t}\Big]
   \label{eqn:equation17}
 \end{aligned}
\end{equation}
\begin{equation}
 \begin{aligned}
\tilde{\beta_\text{t}} := \frac{1-\bar{\alpha}_{\text{t}-1}}{1-\bar{\alpha}_{\text{t}}} \: \beta_\text{t}
   \label{eqn:equation18}
 \end{aligned}
\end{equation}

Then, we derive denoised contrast at each sampling step as follows, 
\begin{equation}
 \begin{aligned}
\tilde{\vec{x}}_{\text{t}-1} = \tilde{\mu_\text{t}}_{avg} + \sqrt{\tilde{\beta_t}} \:\varepsilon ; \: \varepsilon \sim  \mathcal{N}(\varepsilon;0,I).
   \label{eqn:equation19}
 \end{aligned}
\end{equation}

\section{Experiments}

\paragraph{\textbf{Datasets and Baselines.}}

We evaluated our method on two datasets: the BraTS2019 brain tumour dataset \cite{brats} and a healthy dataset  \cite{Islam2023}. From the BraTS2019 dataset, we utilized 305 co-registered multi-contrast MR images, including T1w, T2w, T1CE, and FLAIR. For each scan, we selected 80 middle axial slices, resized them to (256×256), and split the data into 214 subjects for training, 61 for validation, and 30 for testing.

For our healthy dataset, we extracted 100 middle slices from 85 healthy brain MRI scans. We allocated 50, 20 and 15 subjects for training, validation, and testing. In both datasets, one contrast served as the synthesis target, while the remaining contrasts were used as source images to guide the denoising process. 

We compared D2Diff with conventional generative networks, Pix2Pix\cite{Isola2017}, pGAN\cite{Dar2019}, DDPM\cite{Ho2020}, and SOTA MRI synthesis methods including, Hi-Net\cite{Zhou2020}, MM-GAN\cite{Sharma2020}, and SynDiff\cite{ozbey2023} adopted in a supervised manner.


\begin{table*}[t]
\renewcommand{\arraystretch}{1.1}
\resizebox{1\textwidth}{!}{
\begin{tabular}{cccccccccc}
\hline
\textbf{Dataset}                  & \textbf{Contrast}      & \textbf{Metric} & \textbf{pGAN} & \textbf{Pix2Pix} & \textbf{DDPM} & \textbf{MMGAN} & \textbf{Hi-Net} & \textbf{SynDiff} & \textbf{D2Diff}     \\ \hline
\multirow{12}{*}{\textbf{BraTS}}  & \multirow{3}{*}{T1CE}  & PSNR            & 25.13±1.95    & 25.64±1.94       & 24.22±1.81    & 27.72±2.06     & 26.92±2.07      & 28.16±2.36       & \textbf{28.58±2.69} \\ 
                                  &                        & SSIM \%         & 87.34±3.28    & 88.06±3.15       & 56.05±7.37    & 87.47±3.74     & 90.03±2.94      & 91.15±2.88       & \textbf{91.84±2.83} \\ 
                                  &                        & MAE \%          & 3.29±1.43     & 2.93±1.15        & 25.97±5.61    & 11.83±6.83     & 2.66±1.31       & 2.08±1.14        & \textbf{1.97±1.14}  \\ \cline{2-10} 
                                  & \multirow{3}{*}{FLAIR} & PSNR            & 25.37±1.90    & 24.98±1.71       & 25.60±2.03    & 24.50±1.73     & 25.20±1.82      & 27.13±2.11       & \textbf{27.57±2.18} \\ 
                                  &                        & SSIM \%         & 85.69±3.35    & 85.19±3.25       & 76.96±6.57    & 77.57±5.27     & 85.88±3.25      & 89.17±3.50       & \textbf{89.56±3.42} \\
                                  &                        & MAE \%          & 3.32±1.57     & 3.81±1.63        & 25.97±5.61    & 17.58±12.37    & 4.92±2.18       & 2.65±1.24        & \textbf{2.55±1.30}  \\ \cline{2-10} 
                                  & \multirow{3}{*}{T2}    & PSNR            & 25.70±2.01    & 25.52±1.89       & 24.05±1.86    & 26.01±2.12     & 26.26±2.04      & 28.24±2.64       & \textbf{28.73±2.69} \\ 
                                  &                        & SSIM \%         & 89.66±3.80    & 89.27±3.73       & 86.23±4.36    & 87.32±6.40     & 91.49±3.64      & 93.05±4.05       & \textbf{93.51±3.97} \\  
                                  &                        & MAE \%          & 2.50±1.01     & 2.72±1.28        & 3.66±1.20     & 16.93±7.85     & 2.75±1.28       & 1.74±0.95        & \textbf{1.65±0.09}  \\ \cline{2-10} 
                                  & \multirow{3}{*}{T1}    & PSNR            & 26.23±1.89    & 26.71±1.82       & 24.78±1.70    & 25.26±1.92     & 27.19±2.05      & 29.36±2.83       & \textbf{29.96±2.80} \\  
                                  &                        & SSIM \%         & 90.94±3.21    & 91.12±2.96       & 88.18±3.32    & 88.11±4.40     & 93.05±2.83      & 93.63±3.49       & \textbf{94.13±3.33} \\  
                                  &                        & MAE \%          & 2.90±1.80     & 2.46±1.49        & 3.08±1.33     & 8.42±5.91      & 2.42±1.51       & 1.79±1.53        & \textbf{1.71±1.52}  \\ \hline
\multirow{9}{*}{\textbf{Healthy}} & \multirow{3}{*}{FLAIR} & PSNR            & 26.89±1.61    & 26.89±1.61       & 23.13±1.66    & 27.12±1.83     & 28.32±2.09      & 29.30±2.31       & \textbf{29.65±2.25} \\ 
                                  &                        & SSIM \%         & 92.03±2.38    & 91.45±2.94       & 42.77±6.81    & 91.48±3.54     & 94.26±2.15      & 95.01±2.11       & \textbf{95.34±2.02} \\  
                                  &                        & MAE \%          & 10.47±8.05    & 1.81±0.54        & 3.45±0.70     & 6.93±3.47      & 1.50±0.51       & 1.28±0.44        & \textbf{1.25±0.43}  \\ \cline{2-10} 
                                  & \multirow{3}{*}{T2}    & PSNR            & 25.78±1.28    & 24.88±1.56       & 25.19±1.21    & 26.61±1.20     & 27.21±1.20      & 27.70±1.63       & \textbf{28.54±1.67} \\ 
                                  &                        & SSIM \%         & 89.07±3.76    & 87.26±4.80       & 79.24±5.80    & 88.88±4.52     & 92.00±2.95      & 92.72±2.90       & \textbf{93.56±2.66} \\  
                                  &                        & MAE \%          & 25.90±6.70    & 2.23±0.66        & 2.35±0.54     & 16.76±5.96     & 2.08±0.69       & 1.51±0.46        & \textbf{1.38±0.43}  \\ \cline{2-10} 
                                  & \multirow{3}{*}{T1}    & PSNR            & 27.68±1.63    & 26.83±1.89       & 26.26±1.83    & 29.60±1.57     & 29.16±2.00      & 30.09±2.28       & \textbf{30.82±2.33} \\  
                                  &                        & SSIM \%         & 93.49±2.49    & 92.22±3.19       & 86.85±4.00    & 93.84±3.08     & 95.32±2.17      & 95.76±2.03       & \textbf{96.23±1.91} \\  
                                  &                        & MAE \%          & 12.43±1.70    & 1.78±0.56        & 2.09±0.61     & 6.85±4.99      & 1.33±0.52       & 1.19±0.42        & \textbf{1.10±0.39}  \\ \hline
\end{tabular}}
\caption{Performance comparison for healthy and BraTS datasets (mean±std) for different synthesis contrasts. The best performance is in bold with statistical significance $p$ < 0.05 based on paired mean t-test .}
\label{tab:quantitative_results}
\end{table*}

\section{ Experimental Results}

\begin{figure*}[ht]
\includegraphics[width=\textwidth]{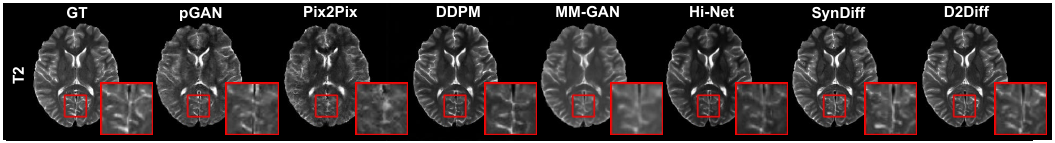}
\caption{Visualization of synthetic MRI results on healthy dataset.
} \label{fig:qualitative_results_healthy}
\end{figure*}

\begin{figure*}[!htb]
\includegraphics[width=\textwidth]{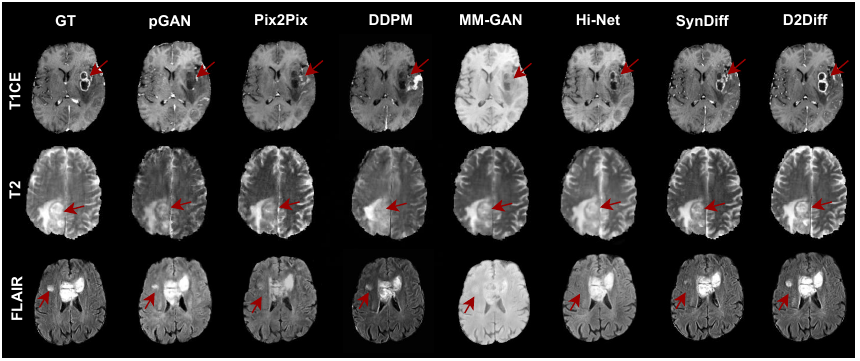}
\caption{Visualization of synthetic MRI results on BraTS dataset.
} \label{fig:qualitative_results_brats}
\end{figure*}

The qualitative performance of D2Diff is illustrated in Fig. \ref{fig:qualitative_results_healthy} and \ref{fig:qualitative_results_brats} for both healthy and BraTS synthetic results. For healthy subjects, D2Diff better preserves anatomical structures, offering superior contrast details compared to other methods. In tumour datasets, it improves lesion synthesis, particularly in challenging tasks like T1CE, where other methods struggle with contrast enhancements. It produces sharper tumour boundaries that closely resemble ground truths. Additionally, the quantitative evaluation in Table \ref{tab:quantitative_results} for PSNR, SSIM, MAE\cite{Dohmen2025} confirms D2Diff's superiority, outperforming all methods across tasks.


\begin{figure*}[ht]
    \begin{minipage}{0.48\textwidth}
        \centering
        \includegraphics[width=\textwidth]{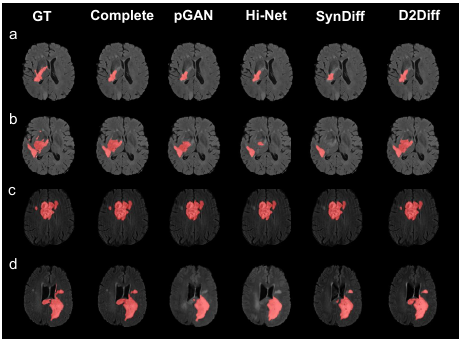}
        \caption{Segmentation results.}
        \label{fig:segmentation}
    \end{minipage}
    \hfill
    \begin{minipage}{0.48\textwidth}
        \centering
        \resizebox{\textwidth}{!}{
            \begin{tabular}{ccccccccc}
            \hline
            \multirow{2}{*}{\textbf{Task}} & \multicolumn{4}{c}{\textbf{Contrasts}} & \multicolumn{4}{c}{\textbf{Dice Score\% ↑}} \\ \cline{2-9} 
                                           & T1     & T2     & FLAIR     & T1CE     & pGAN    & Hi-Net   & SynDiff   & D2Diff   \\ \hline
            a                              & \checkmark      & x      & \checkmark         & \checkmark        & 80.5    & 80.02    & 80.91     & \textbf{81.05}  \\ \hline
            b                              & \checkmark      & x      & \checkmark         & x        & 80.64   & 79.70     & 80.81     & \textbf{81.19}  \\ \hline
            c                              & x      & x      & \checkmark         & x        & 79.96   & 78.93    & 80.66     & \textbf{81.02}  \\ \hline
            d                              & x      & x      & x         & x        & 76.50    & 73.49    & 79.31     & \textbf{79.34}  \\ \hline
             complete                              & \checkmark      & \checkmark      & \checkmark         & \checkmark        & &80.83  \\ \hline
            \end{tabular}
        }
        \captionsetup{font=footnotesize}
         \captionof{table}{\centering Segmentation performance.}
        \label{tab:segmentation_dice_score}

        \vspace{1mm} 

        \resizebox{\textwidth}{!}{
            \begin{tabular}{cccc}
            \hline
            \multicolumn{1}{l}{}       & PSNR ↑                & SSIM\% ↑               & MAE\% ↓                \\ \hline
            Freq. guidance (H1)        & 28.01±2.41          & 91.24±2.88          & 2.09±1.19          \\ \hline
            Spatial guidance (H2)      & 28.39±2.45          & 91.82±2.87          & 2.02±1.30          \\ \hline
            w/o freq. feat. adaptation & 28.54±2.32          & 91.34±2.82          & 2.07±1.27          \\ \hline
            w/o mask loss       & 28.29±2.40          & 91.57±2.80          & 2.04±1.25          \\ \hline
            D2Diff                       & \textbf{28.58±2.69} & \textbf{91.84±2.83} & \textbf{1.97±1.14} \\ \hline
            \end{tabular}
        }
        \captionsetup{font=footnotesize}
        \captionof{table}{\centering Ablation study.}
        \label{tab:ablation}
    \end{minipage}
\end{figure*}

\paragraph{\textbf{Downstream segmentation task performance.}}
To assess the diagnostic equivalence of our synthetic results, we conducted tumour segmentation using the BraTS dataset. A MONAI U-Net\cite{monai} was trained on all four contrasts to predict tumour masks with the same train-test split as synthesis tasks. Table \ref{tab:segmentation_dice_score} marks contrasts replaced by synthetic images during testing, with Dice scores for predicted masks. Fig \ref{fig:segmentation} presents qualitative comparisons. The "Complete" setup represents segmentation using all actual test contrasts without synthetic replacements. For comparison, we selected top-performing methods from each category. Results show that D2Diff closely matches the "Complete" setup, demonstrating clinically reliable and plausible synthesis quality. Notably, D2Diff achieves slightly higher Dice scores, likely due to the dataset's multi-site variability \cite{brats}, including contrast differences and occasional artifacts. D2Diff's diffusion-based dual-domain architecture effectively mitigates these artifacts and handles contrast variations more robustly, leading to improved segmentation performance.

\paragraph{\textbf{Ablation Study.}}
We conducted an ablation study to assess the impact of individual components and the dual-domain mutual learning approach. The T1CE synthesis task was selected to perform ablation as it is one of the most challenging tasks in tumour synthesis. As shown in Table \ref{tab:ablation}, every component contributes to enhancing overall synthesis quality. Notably, the results highlight the effectiveness of mutual learning between the two domain-guided networks, demonstrating superior performance over their individual synthesis outputs.

\section{Conclusion}
In this work, we introduce a dual-domain diffusion framework for multi-contrast MRI synthesis, leveraging frequency and spatial features to capture both intensity variations and spatial differences across contrasts. Experimental results demonstrate that D2Diff outperforms baseline methods, producing more accurate synthetic images. Additionally, superior downstream tumour segmentation highlights the diagnostic value of the synthetic images. Despite these promising results, further research is necessary to validate clinical reliability across multi-centre datasets, diverse clinical settings and imaging protocol variations. Overall, D2Diff offers a promising approach for high-fidelity multi-contrast MRI synthesis, contributing to the efficiency and safety of medical imaging.


\end{document}